\documentclass[aps,prl,twocolumn,showpacs,superscriptaddress]{revtex4-1}
\usepackage{epsfig,amssymb,amsmath,amsfonts}
\usepackage{graphicx}
\usepackage{color}
\usepackage{pdfpages}

\DeclareMathOperator\erfc{erfc}

\begin{document}

\title{Large Deviations in Single File Diffusion}
\author{P. L. Krapivsky}
\affiliation{Physics Department, Boston University, Boston, MA 02215, USA.}
\author{Kirone Mallick}
\author{Tridib Sadhu}
\affiliation{Institut de Physique Th\'eorique, CEA/Saclay, Gif-sur-Yvette Cedex, France.}
\date{\today}
\begin{abstract}
        We apply macroscopic fluctuation theory to study the diffusion of a tracer
        in a one-dimensional interacting particle system 
        with excluded mutual passage, known as single-file diffusion. 
        In the case of Brownian point particles with hard-core repulsion, 
        we derive the cumulant generating function of the tracer position and its large deviation function.  
        In the general case of arbitrary inter-particle interactions, we express the variance of the 
        tracer position in terms of the collective transport properties, viz. the diffusion coefficient and
        the mobility. Our analysis applies both for fluctuating (annealed) and fixed (quenched) initial
	configurations.
\end{abstract}

\pacs{05.40.-a, 83.50.Ha, 87.16.dp, 05.60.Cd}
\maketitle
\begin{figure}
	\centering{\includegraphics[width=0.4\textwidth]{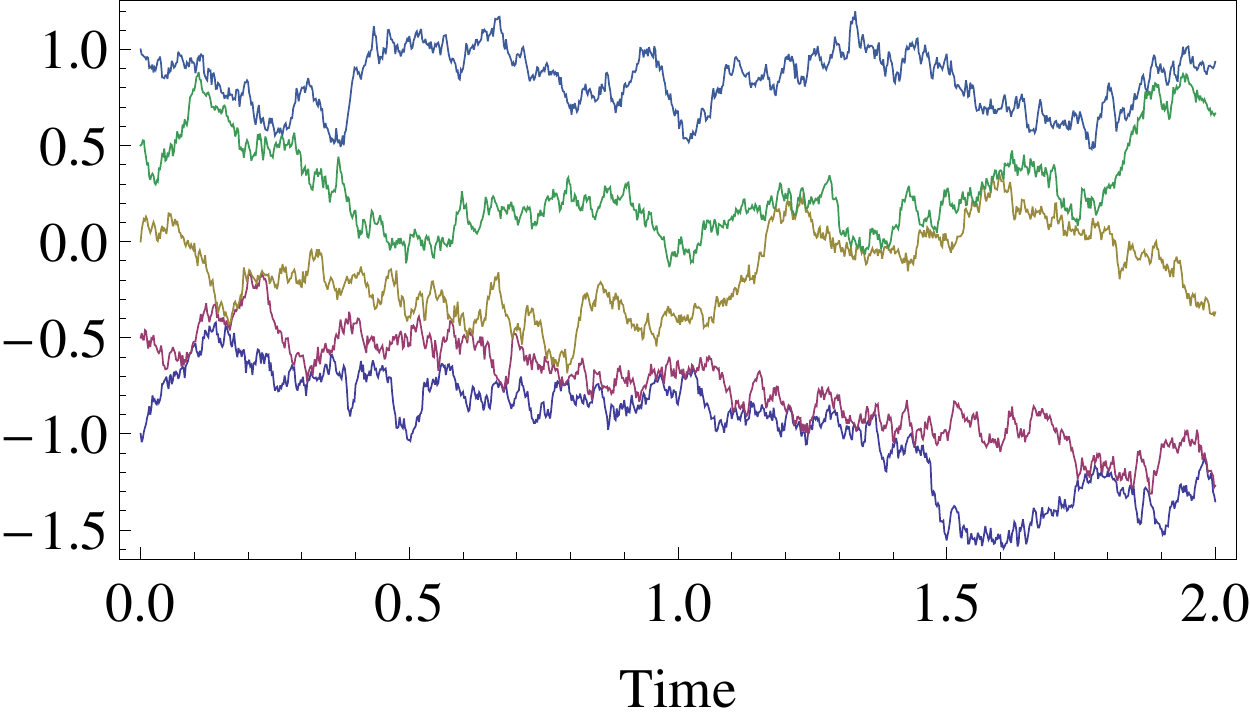}}
\caption{Single-file diffusion of Brownian point particles: individual
trajectories do not cross each other. \label{fig:fig1}}
\end{figure}

Single-file diffusion refers to the motion of interacting diffusing particles in
quasi-one-dimensional channels which are so narrow that particles cannot
overtake each other and hence the order is preserved (see Fig.~\ref{fig:fig1}). Since its introduction
more than 50 years ago to model ion transport through cell
membranes \cite{HODGKIN1955},  single-file diffusion has been observed in a wide
variety of systems, \textit{e.g.}, it describes diffusion of large molecules in
zeolites \cite{KARGER1992,CHOU1999}, transport in narrow pores or in super-ionic
conductors  \cite{Meersmann2000, Richards1977}, and sliding of proteins along DNA \cite{Li2009}.

The key feature of single-file diffusion is that a typical displacement of a tracer particle scales as $t^{1/4}$ rather than $\sqrt{t}$ as in normal diffusion. This  sub-diffusive scaling has been demonstrated in a number of
experimental realizations \cite{KUKLA1996,Wei2000,Lutz2004,Lin2005,Das2010,Siems2012}.
Theoretical analysis leads to a  challenging many-body
problem  \cite{Spohn1991,Ferrari1996}  because the motion of particles is strongly correlated. The sub-diffusive behavior has been explained heuristically for general interactions
 \cite{Alexander1978,Kollmann2003}. Exact results have been mostly established in 
the simplest case of particles with hard-core
repulsion and no other interactions \cite{Harris1965,Levitt1973,Percus1974,Arratia1983,Lizana2008}.

Finer statistical properties of the tracer position, such as  higher cumulants or  the probability distribution 
of rare excursions, require more advanced techniques and they are the main subject of this Letter.
 Rare events are encoded by large deviation functions  \cite{Sethuraman2013}  
that   play  a prominent role in contemporary  developments of statistical physics \cite{TOUCHETTE}. 
Large deviation functions have been computed in a very few cases \cite{Rodenbeck1998,Lizana2008,Gerschenfeld2009,Illien2013} and 
their exact determination in interacting many-particle systems
is a major theoretical challenge \cite{Derrida2011,*Derrida2007}. In single-file systems, the number of particles is usually not too large, and hence large fluctuations can be observable. Recent advances in experimental realizations of single-file systems \cite{Lutz2004,Wei2000,Lin2005,Siems2012,KUKLA1996,Das2010} open the possibility of probing higher cumulants.

The aim of this Letter is to present a systematic approach for calculating the cumulant
generating function of the tracer position in single-file
diffusion. Our analysis is based on macroscopic fluctuation theory, a recently developed 
framework describing dynamical fluctuations in driven diffusive systems (see
\cite{Jona-Lasinio2014,*Jona-Lasinio2010} and references therein). Specifically, we solve the governing equations of macroscopic fluctuation theory in the case of Brownian point particles with hard-core exclusion. This allows us to obtain the cumulants of tracer position and, by a Legendre transform, the large deviation function.

Macroscopic fluctuation theory also provides a simple explanation of 
the long memory effects found in single-file, in which initial conditions
continue to affect the position of the tracer, {\it e.g.}, its variance,
even in the long  time limit \cite{Leibovich2013,Lizana2014}. The statistical properties of the tracer position
are not the same if the initial state is fluctuating or fixed---this situation is akin to 
annealed versus  quenched averaging  in disordered systems \cite{Gerschenfeld2009}.
For general inter-particle interactions, we derive an explicit formula for
the variance of the tracer position in terms of transport coefficients and obtain new results
for the exclusion process. 

We start by formulating the problem of tracer diffusion in terms of macroscopic fluctuation theory, or equivalently fluctuating hydrodynamics. The fluctuating density field  $\rho(x,t)$ satisfies the Langevin equation \cite{Spohn1991} 
\begin{equation}
	\partial_{t}\rho(x,t)=\partial_{x}\left[D(\rho)\partial_{x}\rho(x,t)+\sqrt{\sigma(\rho)}\eta(x,t)\right],
	\label{eq:langevin}
\end{equation}
where $\eta(x,t)$ is a white noise with zero mean and  with variance
$\langle \eta(x,t)\eta(x^{\prime},t^{\prime}) \rangle=\delta(x-x^{\prime})\delta(t-t^{\prime})$. 
The diffusion coefficient $D(\rho)$ and the mobility $\sigma(\rho)$ encapsulate  
the transport characteristics of the diffusive many-particle system, they can be expressed 
in terms of integrated particle current \cite{Bodineau2004}. All the  relevant microscopic details of
inter-particle interactions  are thus embodied,  at the macroscopic scale, 
in these two coefficients.

The position  $X_{T}$  of the tracer particle at time $T$   can be  related to
the fluctuating density field $\rho(x,t)$  by using the single-filing constraint which implies that the total number of particles to  the right of the tracer
does not change with time. Setting the initial tracer position at the origin, we obtain 
\begin{equation}
	\int_0^{X_{T}}\rho(x,T)dx=\int_{0}^{\infty}\left[ \rho(x,T)-\rho(x,0) \right]dx.
	\label{eq:X relate q}
\end{equation}
This relation  defines the tracer's  position  $X_{T}$  as a  functional of the macroscopic 
density field  $\rho(x,t)$.  Variations of  $X_{T}$ smaller than the  coarse-grained
scale are ignored:  their contributions  are  expected  to be  sub-dominant in the limit of a  large time $T$.
The statistics of $X_{T}$ is characterized by  the cumulant generating function 
\begin{equation}
	\mu(\lambda)=\ln\left[\langle \exp(\lambda X_{T})\rangle\right],
	\label{eq:mu}
\end{equation}
where $\lambda$ is a Lagrange multiplier and the angular bracket denotes ensemble average.
We shall calculate  this generating function by using techniques developed by Bertini \textit{et al.} \cite{Bertini2002,*Bertini2001,Jona-Lasinio2014,*Jona-Lasinio2010}, see also \cite{Gerschenfeld2009},
to derive the large deviation function of the  density profile. Starting from 
\eqref{eq:langevin},  the average in \eqref{eq:mu} can be expressed as a path integral
\begin{equation}
	\langle e^{\lambda X_{T}} \rangle=\int \mathcal{D}\left[
	\rho,\hat{\rho}
	\right]e^{-S\left[ \rho,\hat{\rho} \right]},
	\label{eq:path}
\end{equation}
where the action, obtained via the Martin-Siggia-Rose formalism
\cite{Martin1973,DeDominicis1978}, is given by
\begin{eqnarray}
	S[\rho,\hat{\rho}]&&=-\lambda X_{T}+F[\rho(x,0)]+\int_{0}^{T}dt \int_{-\infty}^{\infty} dx
	\nonumber\\
	&&\left[\hat{\rho} \partial_{t}\rho-\tfrac{1}{2}\sigma(\rho)\left(
	\partial_{x}\hat{\rho} \right)^{2}
	+D(\rho)\,\partial_{x}\rho\,\partial_{x}\hat{\rho}\right].
	\label{eq:action}
\end{eqnarray}
Here $F[\rho(x,0)]=-\ln(\text{Prob}[\rho(x,0)])$ and $\hat{\rho}(x,t)$ is the conjugate response field.
We consider two settings, annealed (where we average over initial states drawn from equilibrium) and quenched.
In the annealed case, the large deviation function $F[\rho(x,0)]$ corresponding to the observing of
the density profile $\rho(x,0)$ can be found from the fluctuation dissipation theorem which is satisfied
at equilibrium. This theorem implies \cite{Spohn1991,Derrida2011,*Derrida2007,Krapivsky2012} that $f(r)$, the free energy density of the equilibrium system at density $r$, satisfies $f^{\prime\prime}(r)=2 D(r)/\sigma(r)$. From this one finds \cite{Derrida2011,*Derrida2007,Gerschenfeld2009} 
\begin{equation}
	F[\rho(x,0)]=\int_{-\infty}^{\infty}dx \int_{\rho}^{\rho(x,0)}dr
	\frac{2 D(r)}{\sigma(r)}\left[ \rho(x,0)-r \right],
	\label{eq:F}
\end{equation}
where $\rho$ is the uniform average density at the initial equilibrium state.
In the quenched case, the initial density is fixed, $\rho(x,0)=\rho$, and $F[\rho(x,0)]=0$. 

At large times, the integral in \eqref{eq:path} is dominated by the path
minimizing  the action  \eqref{eq:action}. If  $(q,p)$ denote the functions
$(\rho,\hat{\rho})$ for the optimal action  paths, variational calculus yields two coupled partial differential equations for these optimal paths  
\begin{subequations}
\begin{align}
\label{eq:optimal 1}
	\partial_{t}q-\partial_{x}\left[ D(q)\partial_{x} q\right]&= -
	\partial_{x}\left[\sigma(q)\partial_x p
	\right],\\
 \label{eq:optimal 2}	
	\partial_{t}p+D(q)\partial_{xx}p&= -\tfrac{1}{2}\sigma^{\prime}(q)\left(
	\partial_{x}p
	\right)^{2}. 
\end{align}
\end{subequations}
The boundary conditions are also  found  by  minimizing the  action and   they depend on the initial state  \cite{sup}. In the annealed case,  the boundary conditions read
\begin{eqnarray}
	p(x,T)&=& B\theta(x-Y) \quad\text{with}\quad  B=\lambda/q(Y,T)  \label{eq:annealed pT},\\
	p(x,0)&=& B\theta(x)+\int_{\rho}^{q(x,0)}dr \frac{2 D(r)}{\sigma(r)}\,.
	\label{eq:annealed p0}
\end{eqnarray}
Here $\theta(x)$ is the Heaviside step function, $Y$ is  the value of $X_T$  in 
Eq.~\eqref{eq:X relate q} when the density profile $\rho(x,t)$ is taken to
be  the optimal  profile $q(x,t)$. 
Note that $Y$ representing the tracer position for  the optimal path (at a given value of $\lambda$) 
is a deterministic quantity.  

In the quenched case, the initial configuration is fixed and therefore 
$q(x,0)=\rho$. The `boundary' condition for $p(x,T)$ is the  same as in \eqref{eq:annealed pT}.

In the long  time limit, the cumulant generating function \eqref{eq:mu} is determined by
the minimal action $S[q,p]$. Using Eqs.~\eqref{eq:optimal 1}--\eqref{eq:optimal 2} we obtain
\begin{equation}
	\mu(\lambda) =   \lambda Y
	-F[q]-\int_{0}^{T}dt \int_{-\infty}^{\infty}dx \frac{\sigma(q)}{2}\left(
	\partial_{x}p \right)^{2}.
	\label{eq:gen fnc}
\end{equation}
Thus, the problem of determining the cumulant generating function of the tracer
position has been reduced to solving partial differential equations  for 
$q(x,t)$ and $p(x,t)$ with suitable  boundary conditions.

Two important properties of the single-file diffusion follow from
the  formal solution \eqref{eq:gen fnc}. First, since $\mu(\lambda)$ is an even function of $\lambda$, 
all  odd cumulants of the tracer position vanish. Second, it can be shown that 
$\mu(\lambda)$ is proportional to $\sqrt{T}$: thus, all even cumulants scale as $\sqrt{T}$. If the tracer position $X_T$  is rescaled by $T^{1/4}$, all  cumulants higher than the second 
vanish  when  $T \to \infty$. This leads to the well known result \cite{Arratia1983} that the tracer
position is asymptotically Gaussian.

To determine $\mu(\lambda)$ we need to solve Eqs.~\eqref{eq:optimal
1}--\eqref{eq:optimal 2}. This is impossible for arbitrary  $\sigma(q)$ and
$D(q)$, but for Brownian
particles with hard-core repulsion, where $\sigma(q)=2q$ and $D(q)=1$, an 
exact solution can be found. In the annealed case, Eqs.~\eqref{eq:optimal 1}--\eqref{eq:optimal 2}
for Brownian particles become
\begin{subequations}
\begin{align}
\label{eq:optimal 1sf}
	\partial_{t}q-\partial_{xx}q&=-\partial_{x}\left[ 2 q\partial_x p
	\right],\\
	\partial_{t}p+\partial_{xx}p&=-\left(
	\partial_{x}p
	\right)^{2}. 
\label{eq:optimal 2 sf}
\end{align}
\end{subequations}
The boundary conditions are \eqref{eq:annealed pT} and \eqref{eq:annealed p0}, the latter one simplifies to 
\begin{equation*}
q(x,0)=\rho\, \exp\!\left[p(x,0)-B\theta(x)\right]
\end{equation*}
in the case of Brownian particles. 

We treat $B$ and $Y$ as parameters to  be  determined self-consistently. The
canonical Cole-Hopf transformation from $(q,p)$ to $Q=qe^{-p}$ and $P=e^p$
reduces the non-linear Eqs.~\eqref{eq:optimal 1sf}--\eqref{eq:optimal 2 sf} to
non-coupled linear equations \cite{Elgart2004,Gerschenfeld2009,KMS2012}, a
diffusion equation for $Q$ and an anti-diffusion equation for $P$. Solving
these equations we obtain explicit expressions for $p(x,t)$ and $q(x,t)$
\cite{sup}.

\begin{figure}
	\centering{\includegraphics[width=0.4\textwidth]{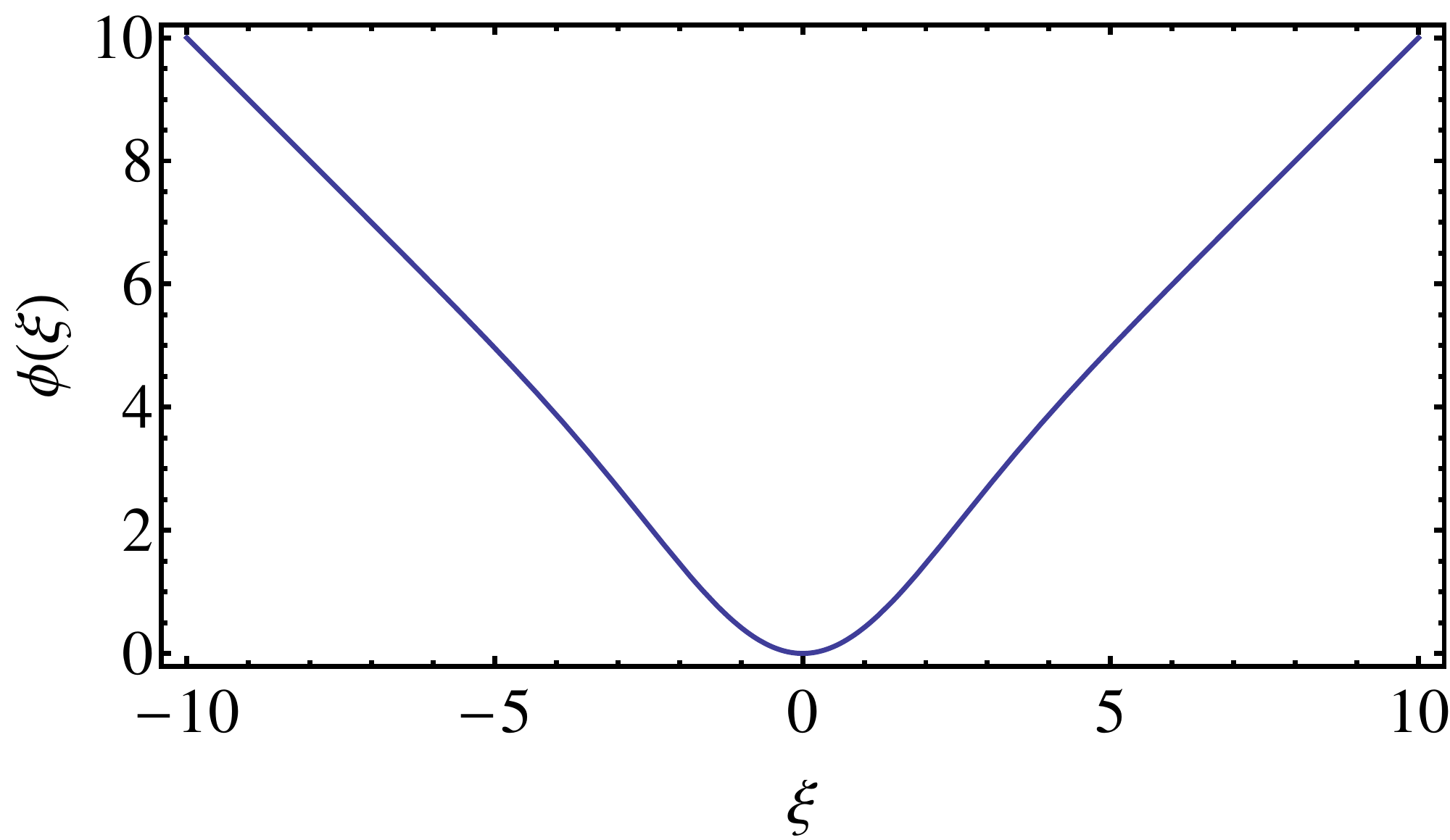}}
\caption{The large deviation function of tracer position
 in the case of  Brownian point particles in the 
annealed setting with density $\rho=1$. \label{fig:fig2} }
\end{figure}

From this solution, the  generating function $\mu(\lambda)$ is obtained 
in a parametric form as
\begin{equation}
	\mu(\lambda)=\left[ \lambda+\rho \frac{1-e^{B}}{1+e^{B}}
	\right]Y,
	\label{eq:mu annealed sf}
\end{equation}
where $Y$ and $B$ are self-consistently  related to $\lambda$ by 
\begin{eqnarray}
	\lambda&=& \rho \left(1- e^{-B} \right)\left[
	1+\tfrac{1}{2}\left(e^{B}-1\right)\erfc(\eta)
	\right]   \\
	e^{2B}&=& 1+\frac{2\eta}{\pi^{-1/2}e^{-\eta^2}-\eta\,\erfc(\eta)}
	\label{lambda_B}
\end{eqnarray}
where we used the shorthand notation $\eta=Y/\sqrt{4T}$. 

The cumulants of the tracer position can be extracted from this parametric
solution by expanding $\mu(\lambda)$ in powers of $\lambda$. 
The first three non-vanishing cumulants are
\begin{subequations}
\begin{align}
	\langle X_{T}^{2} \rangle_{c}&= \frac{2}{\rho
	\sqrt{\pi}}\sqrt{T}, \label{eq:variance annealed}\\
	\langle X_{T}^{4} \rangle_{c}&= \frac{6\left( 4-\pi
	\right)}{\left(\rho \sqrt{\pi}\right)^{3}}\sqrt{T}
	\label{eq:y4 annealed}\\
	\langle X_{T}^{6} \rangle_{c}&= \frac{30\left( 68-30\pi+3\pi^2
	\right)}{\left(\rho \sqrt{\pi}\right)^{5}}\sqrt{T}
	\label{eq:y6 annealed}
\end{align}
\end{subequations}
in the large time limit. 
The expression \eqref{eq:variance annealed} for the variance matches the well-known result \cite{Harris1965,Spohn1991,Leibovich2013}. The exact solution \eqref{eq:mu annealed sf}--\eqref{lambda_B}, which encapsulates \eqref{eq:variance annealed}--\eqref{eq:y6 annealed} and all higher cumulants, is one of our main results. 

The large deviation function of the tracer position,
 defined, in the limit $T \to \infty$,  via
\begin{equation}
	\text{Prob}\left( \frac{X_{T}}{\sqrt{T}}=\xi \right)\sim \exp\left[
	-\sqrt{T}\phi(\xi)\right],\nonumber
\end{equation}
is the Legendre transform of $\mu(\lambda)$, given by the 
parametric solution  \eqref{eq:mu annealed sf}--\eqref{lambda_B}. This large deviation function
$\phi(\xi)$ can be expressed as
\begin{equation}
	\phi(\xi)=\rho\left[ \sqrt{\alpha(\xi)}-\sqrt{\alpha(-\xi)}
	\right]^{2},
	\label{eq:ldf}
\end{equation}
with $\alpha(\xi)=\int_{\xi/2}^{\infty}dz\erfc(z)$. The large deviation function
$\phi(\xi)$ is plotted on
Fig.~\ref{fig:fig2}. The asymptotic formula $\phi(\xi)\simeq \rho |\xi|$ is formally valid when $|\xi|\to\infty$,
but it actually provides an excellent approximation everywhere apart from small $\xi$. The expression \eqref{eq:ldf} matches an
exact microscopic calculation \cite{Rodenbeck1998,Abhishek2014}. 

We carried out a similar analysis for a quenched initial condition.
Here, we cite a few concrete results. The first two even cumulants read 
\begin{subequations}
\begin{align}
	\langle X_{T}^{2}\rangle_{c}
	=&\frac{\sqrt{2}}{\rho\sqrt{\pi}}\sqrt{T},\label{eq:variance quenched}\\
	\langle X_{T}^{4}\rangle_{c}
	=&\frac{2\sqrt{2}}{\rho^{3}\sqrt{\pi}}\left[
	\frac{9}{\pi}\arctan \left( \frac{1}{2\sqrt{2}}\right)-1
	\right]\sqrt{T}.\label{eq:y4 quenched}
\end{align}
\end{subequations}
These cumulants are different from the annealed case. In particular, the variance is 
$\sqrt{2}$ times smaller, in agreement with previous findings \cite{LizanaPRE2010,Krapivsky2012,Leibovich2013}.
An asymptotic analysis yields $\phi(\xi)\simeq \rho |\xi|^3/12$ when $ |\xi| \to \infty$. This asymptotic behavior can also
 be extracted from the knowledge of extreme current fluctuations \cite{Baruch2014}.

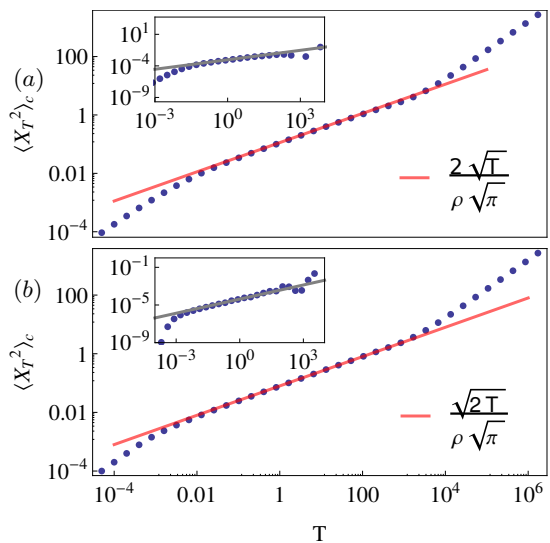
\begin{figure}
	\centering{\includegraphics[width=0.4\textwidth]{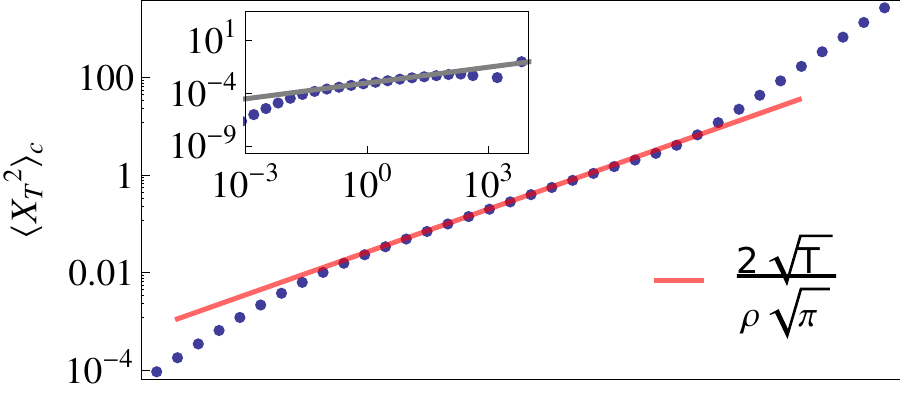}}
	\centering{\includegraphics[width=0.4\textwidth]{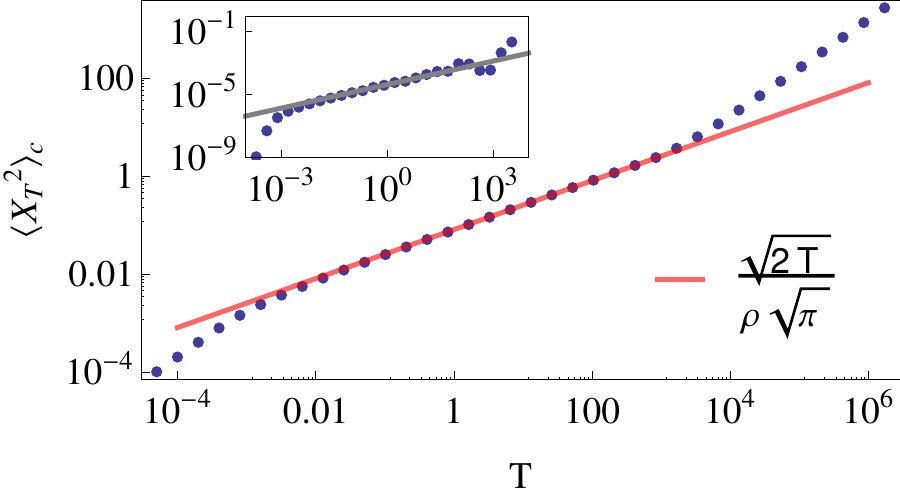}}
\caption{Simulation results for the second cumulant (main plots) and the fourth cumulant (insets). Shown are results for  
Brownian point particles with average density $\rho=10$ in (a) annealed and (b) quenched settings. 
The solid lines denote corresponding theoretical results; the variance
was already computed in \cite{Leibovich2013}.   \label{fig:fig3}}
	\begin{picture}(0,0)
		\put(-100,260){$(a)$}
		\put(-100,180){$(b)$}
	\end{picture}
\end{figure}

To test our predictions, we performed Monte Carlo simulations of single-file diffusion of Brownian point particles.
In most simulations, we considered $2001$ particles on an infinite line which are initially distributed on the interval $[-100,100]$. 
In the annealed case, the particles were distributed randomly; in the quenched case,  they
were uniformly  spaced. The central particle is
the tracer. The cumulants of the tracer  position at different times, determined by
averaging over $10^{8}$ samples are shown in Fig.~\ref{fig:fig3}. At small times
(comparable to the mean collision time), the tracer diffusion is
normal. At very long times, the diffusion again becomes
normal since there is only a finite number of particles in our simulations. The
crossover time to normal diffusion increases as $N^2$ with the number of
particles. At intermediate times, the
motion is sub-diffusive and the cumulants scale as $\sqrt{T}$. In this range the data
are in excellent agreement with theoretical predictions \eqref{eq:variance
annealed}--\eqref{eq:y4 annealed} and  \eqref{eq:variance
quenched}--\eqref{eq:y4 quenched}.

For arbitrary $\sigma(\rho)$ and $D(\rho)$, the governing equations \eqref{eq:optimal
1}--\eqref{eq:optimal 2} are intractable, so one has to resort to numerical methods \cite{Bunin2012,Krapivsky2012}. 
For small values of $\lambda$, however, a perturbative expansion of $p(x,t)$ and $q(x,t)$ with respect to  $\lambda$
can be performed \cite{Krapivsky2012}. This is feasible because for $\lambda=0$ the
solution is $p(x,t)=0$ and $q(x,t)=\rho$, for both types  of initial conditions.
Equations \eqref{eq:optimal 1}--\eqref{eq:optimal 2} give rise to  a hierarchy 
 of diffusion equations with source terms.  For example, to the linear
order  in  $\lambda$, we have 
\begin{eqnarray*}
	\partial_{t}p_{1}+D(\rho)\partial_{xx}p_{1}&=&0,\\
	\partial_{t}q_{1}-D(\rho)\partial_{xx}q_{1}&=&-\sigma(\rho)\partial_{xx}p_{1},
\end{eqnarray*}
where $p_{1}$ and $q_{1}$ are the first  order  terms  in the expansions of
$p$ and $q$, respectively. Solving above equations and noting that 
$\langle X_{T}^{2}\rangle_{c}$ is a function of the $p_1$
and $q_1$, we obtain a general formula for the variance \cite{sup}
\begin{equation}
	\langle X_{T}^{2} \rangle_{c} = 
	\frac{\sigma(\rho)}{\rho^{2}\sqrt{\pi}}\sqrt{\frac{T}{D(\rho)}}
	\label{eq:annealed w} 
\end{equation}
in the annealed case. In the quenched case, the variance is given by the same expression but with an additional $\sqrt{2}$ term in the denominator. We emphasize that Eq.~\eqref{eq:annealed w}
 applies to general single-file systems, ranging from hard-rods
\cite{Alexander1978} to colloidal suspensions \cite{Kollmann2003}, and also to
lattice gases \cite{Arratia1983}. As an example of the latter, consider the symmetric simple exclusion
process (SEP). For this lattice gas, the transport coefficients are $D(\rho)=1$ and $\sigma(\rho)=2\rho(1-\rho)$
(we measure length in the unit of lattice spacing, so $0<\rho<1$ due to the exclusion condition),
so Eq.~\eqref{eq:annealed w}  yields
$\langle X^{2}_{T}\rangle_{c}=2(1-\rho)\sqrt{T}/\rho\sqrt{\pi}$,
in agreement with well-known results \cite{Arratia1983}. The result for
colloidal suspension derived in \cite{Kollmann2003} is recovered  by inserting 
 in \eqref{eq:annealed w} 
the fluctuation dissipation relation $\sigma(\rho)=2 S(\rho) D(\rho)$, where  $S(\rho)$ is
the structure factor \cite{Spohn1991}.

Finding  higher cumulants from  the perturbative expansion leads to 
tedious calculations.  For the SEP,  we have computed the fourth cumulant
 \begin{subequations}
 \begin{align}
	 \langle X_{T}^{4}\rangle_{c}
	 &=\frac{2}{\sqrt{\pi}}\,\frac{1-\rho}{\rho^{3}}\,a(\rho)\sqrt{T}\,,\nonumber\\
	 a(\rho) &= 1-\left( 4-(8-3\sqrt{2})\rho \right)(1-\rho) +\frac{12}{\pi}(1-\rho)^{2}. \nonumber
 \end{align}
 \end{subequations}
in the annealed case. 
For small  values of $\rho$, the above results reduces to \eqref{eq:y4
annealed}.  The complete calculation of the tracer's large deviation function
for the SEP remains a very challenging open problem.

To conclude, we analyzed single-file diffusion employing the
macroscopic fluctuation theory.  For Brownian point particles with
hard-core exclusion, we calculated the full statistics of tracer's position, viz. we derived
an exact  parametric representation for the
cumulant generating function.
We extracted explicit formulas for the first few cumulants and obtained large deviation functions.
We also derived the sub-diffusive scaling of  the  cumulants
and the closed expression \eqref{eq:annealed w} for  the variance, valid for general single-file processes. All our results have been derived in the equilibrium situation (homogeneous initial conditions). It seems possible to extend our approach to non-equilibrium settings. Another interesting direction is to analyze a tracer in an external potential
\cite{Illien2013,Barkai2009,Burlatsky1996,Burlatsky1992} and biased  diffusion
\cite{Imamura2007,Majumdar1991}.

We thank S. Majumdar,  S. Prolhac, T. Sasamoto and R. Voituriez
for fruitful discussions, and A. Dhar for sharing results prior to publication. We are grateful to B. Meerson for 
crucial remarks and suggestions. We thank the Galileo Galilei Institute for
Theoretical Physics for excellent working conditions and the INFN for partial support during
the completion of this work.
\bibliography{reference}

\clearpage
\newpage

\begin{widetext}
	\begin{center}
		

\section*{Supplementary material}

# Supplemental note for the *Letter* "Large deviations in single-file diffusion"

P. L. Krapivsky[1], Kirone Mallick[2] and Tridib Sadhu[2]

[1] Physics Department, Boston University, Boston, MA 02215, USA.
[2] Institut de Physique Théorique, CEA/Saclay, Gif-sur-Yvette Cedex, France.

**Abstract.** In this supplemental note, we present a derivation of the variational formulation of the single-file diffusion. Subsequently, we present an explicit solution of the variational problem for the single-file diffusion of the Brownian point particles. Lastly, we present a derivation of the expression for the variance of tracer position in a general single-file diffusion.

## 1. Variational formulation of the single-file diffusion

As discussed in the *Letter*, the time evolution of the hydrodynamic density profile $\rho(x,t)$ in an one-dimensional single-file diffusion is governed by

$$\partial_t \rho = \partial_x \left[ D(\rho)\partial_x \rho + \sqrt{\sigma(\rho)} \eta \right], \tag{1}$$

where $\eta(x,t)$ is a Gaussian noise with mean zero and covariance

$$\langle \eta(x,t)\eta(x',t') \rangle = \delta(x-x')\delta(t-t'). \tag{2}$$

The angular bracket denotes ensemble average. The $D(\rho)$ and $\sigma(\rho)$ are the diffusivity and mobility, respectively.

The tracer position $X_T$ is a random variable which depends on the particular history of $\rho(x,t)$. The statistics of $X_T$ can be encoded in its moment generating function

$$\langle e^{\lambda X_T} \rangle = 1 + \lambda \langle X_T \rangle + \frac{\lambda^2}{2} \langle X_T^2 \rangle + \cdots,$$

where $\langle X_T^n \rangle$ is the $n$th moment. One can also use the cumulant generating function given by

$$\log \langle e^{\lambda X_T} \rangle = \lambda \langle X_T \rangle_c + \frac{\lambda^2}{2} \langle X_T^2 \rangle_c + \cdots \tag{3}$$

For example, the second cumulant is equal to the variance, $\langle X_T^2 \rangle_c = \langle X_T^2 \rangle - \langle X_T \rangle^2$.

For the tagged particle in single-file diffusion, the $2n$-th moment scales according to $\langle X_T^{2n} \rangle \sim \langle X_T^2 \rangle^n \sim T^{n/2}$. The cumulants which capture the sub-leading corrections scale as $\sqrt{T}$ i.e., $\langle X_T^{2n} \rangle_c \sim \sqrt{T}$.

Considering all possible evolution of $\rho(x,t)$ in the time interval $[0,T]$, the moment generating function of $X_T$ can be written as a path integral

$$\langle e^{\lambda X_T} \rangle = \int \mathcal{D}[\rho] e^{\lambda X_T[\rho] - F[\rho(x,0)]} \left\langle \delta\left(\partial_t \rho - \partial_x \left[D(\rho)\partial_x \rho + \sqrt{\sigma(\rho)}\,\eta\right]\right) \right\rangle_\eta, \quad (4)$$

where $\lambda$ is a parameter (essentially a Lagrange multiplier) and the subscript $\eta$ in the angular bracket on the right-hand side denotes the averaging over history of the noise. Further, $F[\rho(x,0)]$ is a shorthand notation for $F[\rho(x,0)] = -\ln(\text{Prob}[\rho(\text{x},0)])$, where $\text{Prob}[\rho(\text{x},0)]$ is the probability of a density profile $\rho(x,0)$ in the initial state. We follow the standard notation where the square bracket denotes a functional.

The noise average of the Dirac delta function $\delta(y)$ can be replaced by a path integral over a conjugate field $\hat{\rho}(x,t)$, which following the Martin-Siggia-Rose formalism (Ref [33,34] in the *Letter*) leads to

$$\langle e^{\lambda X_T} \rangle = \int \mathcal{D}[\rho, \hat{\rho}] e^{-S[\rho,\hat{\rho}]}, \quad (5)$$

where the action

$$S[\rho, \hat{\rho}] = -\lambda X_T[\rho] + F[\rho(x,0)]$$
$$+ \int_0^T dt \int_{-\infty}^{\infty} dx \left\{ \hat{\rho}\partial_t \rho - \frac{\sigma(\rho)}{2}(\partial_x \hat{\rho})^2 + D(\rho)(\partial_x \rho)(\partial_x \hat{\rho}) \right\}.$$

The action $S[\rho, \hat{\rho}]$ is extensive in time and at large $T$ the path integral is dominated by the path that corresponds to the least action. Let $(q,p) = (\rho, \hat{\rho})$ for this optimal path. To determine this path, we perform a small variation $\rho \to q + \delta\rho$ and $\hat{\rho} \to p + \delta\hat{\rho}$ around the $(q,p)$. The change in action $S[\rho, \hat{\rho}]$ corresponding to this variation, is

$$\delta S = \int_{-\infty}^{\infty} dx \left\{ -\lambda \frac{\delta X_T[q]}{\delta q(x,0)} + \frac{\delta F[q(x,0)]}{\delta q(x,0)} - p(x,0) \right\} \delta\rho(x,0)$$
$$+ \int_{-\infty}^{\infty} dx \left\{ -\lambda \frac{\delta X_T[q]}{\delta q(x,T)} + p(x,T) \right\} \delta\rho(x,T)$$
$$+ \int_0^T dt \int_{-\infty}^{\infty} dx \left\{ -\partial_t p - \frac{\sigma'(q)}{2}(\partial_x p)^2 - D(q)\partial_{xx} p \right\} \delta\rho(x,t)$$
$$+ \int_0^T dt \int_{-\infty}^{\infty} dx \left\{ \partial_t q + \partial_x(\sigma(q)\partial_x p) - \partial_x(D(q)\partial_x q) \right\} \delta\hat{\rho}(x,t), \quad (6)$$

where the functional derivatives are taken at the optimal path $(q,p)$. In the above we used that, $X_T[\rho]$ is a functional of only the initial $\rho(x,0)$ and the final $\rho(x,T)$ density profiles, related by the definition

$$\int_0^{X_T} \rho(x,T) dx = \int_0^{\infty} [\rho(x,T) - \rho(x,0)]\, dx, \quad (7)$$

where it is assumed that the tracer starts at the origin at $t = 0$. As discussed in the *Letter*, this relation is a consequence of the single-filing constraint.

For the action $S[q,p]$ to be minimum, the variation $\delta S$ must vanish, $\delta S[q,p] = 0$. As, in general, $\delta\rho(x,t)$ and $\delta\hat{\rho}(x,t)$ are non-zero, their coefficients in (6) must vanish,

which leads to the following governing equations

$$\partial_t q - \partial_x (D(q)\partial_x q) = -\partial_x (\sigma(q)\partial_x p), \tag{8}$$

$$\partial_t p + D(q)\partial_{xx} p = -\frac{1}{2} \sigma'(q) (\partial_x p)^2. \tag{9}$$

The boundary conditions come from vanishing of the first two integrals in (6). We examine two types of initial states, annealed and quenched. In the first setting, where the initial state is in equilibrium, both $\rho(x,0)$ and $\rho(x,T)$ are fluctuating, equivalently, $\delta q(x,0)$ and $\delta q(x,T)$ are non-zero. Then, for the minimal action, their coefficients in (6) must vanish, leading to the boundary condition

$$p(x,0) = -\lambda \frac{\delta X_T[q]}{\delta q(x,0)} + \frac{\delta F[q(x,0)]}{\delta q(x,0)}, \tag{10}$$

$$p(x,T) = \lambda \frac{\delta X_T[q]}{\delta q(x,T)}. \tag{11}$$

Using the equilibrium distribution in the initial state, it can be shown (Ref [25] of the *Letter*) that,

$$F[q(x,0)] = \int_{-\infty}^{\infty} dx \int_{\rho}^{q(x,0)} dr \frac{2D(r)}{\sigma(r)} [q(x,0) - r], \tag{12}$$

where $\rho$ is the uniform average density at the initial state. Using this expression and the relation in (7), the boundary conditions (10)–(11) transform to

$$p(x,0) = B\theta(x) + \int_{\rho}^{q(x,0)} dr \frac{2D(r)}{\sigma(r)}, \tag{13}$$

$$p(x,T) = B\theta(x-Y), \qquad B = B(\lambda, Y, T) \equiv \lambda/q(Y,T). \tag{14}$$

Here $\theta(x)$ is the Heaviside step function and $Y$ denotes the value of $X_T$ corresponding to the optimal path.

For the quenched case, the initial configuration is fixed at a uniform profile with density $\rho$. This leads to $\delta\rho(x,0) = 0$, and then, the first integral in (6) vanish. The condition for the second integral to vanish yields the same boundary condition as in (14). Thus the boundary conditions for the quenched initial state are

$$q(x,0) = \rho, \tag{15}$$

$$p(x,T) = B\theta(x-Y). \tag{16}$$

For both type of initial states, the cumulant generating function $\mu(\lambda) = \ln\left(\langle e^{\lambda X_T}\rangle\right)$ is determined by the least action $S[q,p]$. Using Eqs. (8)–(9) in the expression (6) we deduce the cumulant generating function

$$\mu(\lambda) = -S[q,p] = \lambda Y - F[q(x,0)] - \int_0^T dt \int_{-\infty}^{\infty} dx \frac{\sigma(q(x,t))}{2} (\partial_x p(x,t))^2. \tag{17}$$

In establishing (17) we have taken into account that the derivatives $\partial_x p$ and $\partial_x q$ vanish as $x \to \pm\infty$.

## 2. Brownian point particles

We now present a detailed solution for Brownian point particles with hard-core repulsion. In this case $\sigma(\rho) = 2\rho$ and $D(\rho) = 1$, so Eqs. (8)–(9) become

$$\partial_t q - \partial_{xx} q = -\partial_x (2q\partial_x p), \tag{18}$$
$$\partial_t p + \partial_{xx} p = -(\partial_x p)^2. \tag{19}$$

Before we consider the specific boundary conditions for the two initial states, we note that the governing equations admit a great simplification. Using a canonical Cole-Hopf transformation, namely writing

$$p = \ln P \quad \text{and} \quad q = Qe^p = QP, \tag{20}$$

one transforms (18)–(19) into linear diffusion equations

$$\partial_t P + \partial_{xx} P = 0, \quad \partial_t Q - \partial_{xx} Q = 0.$$

Solving these equations and using the ansatz (20) we arrive at a formal solution for $p(x,t)$ and $q(x,t)$:

$$p(x,t) = \ln\left[\int_{-\infty}^{\infty} dz\, e^{p(z,T)} \frac{e^{-\frac{(z-x)^2}{4(T-t)}}}{\sqrt{4\pi(T-t)}}\right], \tag{21}$$

$$q(x,t) = e^{p(x,t)} \int_{-\infty}^{\infty} dz\, q(z,0)\, e^{-p(z,0)} \frac{e^{-\frac{(z-x)^2}{4t}}}{\sqrt{4\pi t}}. \tag{22}$$

In the following, we use this solution in the annealed and quenched cases and determine the corresponding cumulant generating function.

### 2.1. Annealed setting

For Brownian particles, $\sigma(q) = 2q$ and $D = 1$. The boundary conditions (13)–(14) become

$$q(x,0) = \rho \exp(p(x,0) - B\theta(x)), \quad p(x,T) = B\theta(x - Y), \tag{23}$$

In determining the solution of the governing equations we first treat both $B$ and $Y$ as parameters and later determine them self-consistently.

Substituting the boundary conditions in the formal solution (21)–(22) and completing the integrals yield

$$p(x,t) = \log\left[1 + (e^B - 1)\frac{1}{2}\mathrm{erfc}\left(\frac{Y-x}{2\sqrt{T-t}}\right)\right], \tag{24}$$

and

$$q(x,t) = \rho\left[1 + (e^{-B} - 1)\frac{1}{2}\mathrm{erfc}\left(\frac{x-Y}{2\sqrt{T-t}}\right)\right]\left[1 + (e^B - 1)\frac{1}{2}\mathrm{erfc}\left(\frac{x}{2\sqrt{t}}\right)\right], \tag{25}$$

where $\mathrm{erfc}(x)$ is the complementary error function.

The goal is now to express $B$ and $Y$ in terms of $\lambda$. The relation between $B$ and $Y$ comes from using the solution in the definition of $Y$,

$$\int_0^Y q(x,T)dx = \int_0^\infty dz\, [q(x,T) - q(x,0)], \tag{26}$$

which is Eq. (7) with $X_T$ and $\rho(x,t)$ replaced by their optimal values. Combining the above relation with (24)–(25) we establish:

$$e^{2B} = 1 + \frac{Y}{2\sqrt{T}}\left[\frac{e^{-\frac{Y^2}{4T}}}{\sqrt{4\pi}} - \frac{Y}{4\sqrt{T}}\mathrm{erfc}\left(\frac{Y}{2\sqrt{T}}\right)\right]^{-1}, \tag{27}$$

which is Eq. (14) in the *Letter*.

To determine $\mu(\lambda)$ we first write it as

$$\mu(\lambda) = \lambda Y - \int_{-\infty}^\infty dx \left(q(x,0)\ln\left(\frac{q(x,0)}{\rho}\right) - (q(x,0)-\rho)\right)$$
$$- \int_0^T dt \int_{-\infty}^\infty dx\, q(x,t)(\partial_x p(x,t))^2, \tag{28}$$

where we have used $\sigma(q) = 2q$, $D(q) = 1$, and $F[q(x,0)]$ from (12). To simplify (28) we use the identity

$$q(\partial_x p)^2 = \partial_t(qp) - \partial_x(p\partial_x q - q\partial_x p - 2pq\partial_x p), \tag{29}$$

which follows from Eqs. (18)–(19). Using above identity together with initial conditions (23) we deduce an expression

$$\mu(\lambda) = \lambda Y + \int_{-\infty}^\infty dx\, (q(x,0) - \rho). \tag{30}$$

Further simplification comes from using the solution $q(x,t)$ in (25) and the relation (27), which yields

$$\mu(\lambda) = \left(\lambda + \rho\frac{1-e^B}{1+e^B}\right)Y, \tag{31}$$

which is Eq. (12) in the *Letter*.

We now derive the second relation between $\lambda$, $B$ and $Y$. This is found by optimizing the value of $\mu(\lambda)$ with respect to the parameter $B$. The derivative must vanish: $\frac{d\mu(\lambda)}{dB} = 0$. It is easier to use equivalently

$$\frac{d\mu(\lambda)}{dY} = 0. \tag{32}$$

This along with (31) yields

$$\lambda = \rho\left(1 - e^{-B}\right)\left[1 + \left(e^B - 1\right)\frac{1}{2}\mathrm{erfc}\left(\frac{Y}{2\sqrt{T}}\right)\right], \tag{33}$$

which is Eq. (13) in the *Letter*.

The expression for $\mu(\lambda)$ in (31) along with the relation (33) and (27) constitutes a parametric solution of the cumulant generating function, reported in the *Letter*.

*2.2. Quenched initial state*

The boundary conditions (15)–(16) become

$$q(x,0) = \rho \quad \text{and} \quad p(x,T) = B\theta(x-Y), \tag{34}$$

where $B$ is defined in (14). Since the optimal equations for $p(x,t)$ and its boundary condition are same in the annealed and the quenched case, the solution is also same and given in (24). On the other hand, the boundary condition (34) for $q(x,t)$ is different, which using the formal solution (22) yields

$$q(x,t) = \rho e^{p(x,t)} \int_{-\infty}^{\infty} dz \frac{1}{1+(e^B-1)\frac{1}{2}\mathrm{erfc}\left(\frac{Y-z}{2\sqrt{T}}\right)} \left( \frac{e^{-\frac{(z-x)^2}{4t}}}{\sqrt{4\pi t}} \right). \tag{35}$$

Similar to the annealed case, we determine $Y$ using the relation (26) together with the solution $p(x,t)$ and $q(x,t)$, viz. (24) and (35), yielding

$$\frac{Y}{2\sqrt{T}} = \int_{-\infty}^{\infty} dz \left[ \frac{(e^B-1)\frac{1}{2}\mathrm{erfc}(-z)}{1+(e^B-1)\frac{1}{2}\mathrm{erfc}(-z)} \right] \frac{1}{2}\mathrm{erfc}(z), \tag{36}$$

The expression for $\mu(\lambda)$ can be obtained by using $\sigma(q) = 2q$ in (17) which becomes

$$\mu(\lambda) = \lambda Y - \int_0^T dt \int_{-\infty}^{\infty} dx\, q\, [\partial_x p]^2, \tag{37}$$

since $F[q(x,0)] = 0$ in the quenched case. The expression further simplifies using the identity in (29) yielding

$$\mu(\lambda) = \lambda Y - \int_{-\infty}^{\infty} dx\, q(x,T)p(x,T) + \int_{-\infty}^{\infty} dx\, q(x,0)p(x,0). \tag{38}$$

Using the solution for $p(x,t)$ and $q(x,t)$, the expression becomes

$$\mu(\lambda) = \lambda Y + \sqrt{4DT}\rho \int_{-\infty}^{\infty} dz \left[ \ln\left\{ 1 + \left(e^B-1\right)\frac{1}{2}\mathrm{erfc}(-z)\right\} \right.$$
$$\left. -Be^B \frac{\frac{1}{2}\mathrm{erfc}(-z)}{1+(e^B-1)\frac{1}{2}\mathrm{erfc}(-z)} \right] \tag{39}$$

Finally, the relation between $B$, $Y$ and $\lambda$ can be obtained by minimizing the above expression for $\mu(\lambda)$ with respect to $B$ i.e., $\frac{d\mu(\lambda)}{dB} = 0$. The emerging relation between $\lambda$ and $B$ does not have a simple form, but one can use it to extract the second and the fourth cumulant presented in the paper.

## 3. The variance of $X_T$ for general $\sigma(q)$ and $D(q)$

The variance of the tracer position can be determined for arbitrary $\sigma(q)$ and $D(q)$ using a perturbative expansion for small $\lambda$. We write an expansion

$$q(x,t) = \rho + \lambda q_1(x,t) + \lambda^2 q_2(x,t) + \cdots, \tag{40}$$
$$p(x,t) = \lambda p_1(x,t) + \lambda^2 p_2(x,t) + \cdots, \tag{41}$$

where we have used that, for $\lambda = 0$, as the evolution is not constrained by the observable $X_T$, so the profile is a hydrodynamic solution $q(x,t) = \rho$ and $p(x,t) = 0$ with $\rho$ being

the average initial density. This can also be justified by using Eqs. (9)–(8) and the boundary conditions for either type of initial states: quenched and annealed.

With this perturbative expansion, we obtain

$$\partial_t p_1 + D(\rho)\partial_{xx} p_1 = 0, \tag{42}$$

$$\partial_t q_1 - D(\rho)\partial_{xx} q_1 = -\sigma(\rho)\partial_{xx} p_1 \tag{43}$$

in the linear order in $\lambda$. From (7) we get $Y_0 = 0$, while the linear order term is

$$Y_1 = \frac{1}{\rho}\int_0^\infty dx \left[q_1(x,T) - q_1(x,0)\right]. \tag{44}$$

These are the only quantities required for determining the variance of $X_T$, which by definition is the order $\lambda^2$ term in $\mu(\lambda)$. Using the perturbative expansion in the expression (17) we get

$$\frac{1}{2!}\langle X_T^2\rangle = Y_1 - F_2 - \frac{\sigma(\rho)}{2}\int_0^T dt \int_{-\infty}^\infty dx\, (\partial_x p_1)^2, \tag{45}$$

where $F_2$ is the order $\lambda^2$ term of the quantity $F[q(x,0)]$, which depends on the initial state. For the quenched initial state $F[q(x,0)] = 0$ and hence $F_2 = 0$. For the annealed initial state we use Eq. (12) and find

$$F_2^{\text{annealed}} = \frac{D(\rho)}{\sigma(\rho)}\int_{-\infty}^\infty dx\, (q_1(x,0))^2. \tag{46}$$

We now determine the variance $\langle X_T^2\rangle$ by solving Eqs. (42)–(43) with the specific boundary conditions for the quenched and the annealed initial states.

*3.1. Quenched initial state*

Plugging the perturbative expansion into (15)–(16) yields

$$p_1(x,T) = \rho^{-1}\theta(x) \quad \text{and} \quad q_1(x,0) = 0 \tag{47}$$

in the linear order. The solution of (42) with the above boundary condition satisfies

$$\partial_x p_1(x,t) = \rho^{-1} R(x,t|0,T), \tag{48}$$

where

$$R(x,t|z,T) = \frac{1}{\sqrt{4\pi D(\rho)(T-t)}} \exp\left[-\frac{(x-z)^2}{4D(\rho)(T-t)}\right] \tag{49}$$

for $0 \leq t \leq T$. Since $p_1(x,t) = 0$ at $x \to -\infty$ we get

$$p_1(x,t) = -\frac{1}{\rho}\int_{-\infty}^x dz\, R(z,t|0,T) = \frac{1}{2\rho}\mathrm{erfc}\left(\frac{-x}{2\sqrt{D(\rho)(T-t)}}\right). \tag{50}$$

The solution for $q_1(x,t)$ in (43) can be written as $q_1(x,t) = -\partial_x \psi(x,t)$ with

$$\psi(x,t) = \frac{\sigma(\rho)}{\rho}\int_0^t d\tau \int_{-\infty}^\infty dz\, G(x,t|z,\tau) R(z,\tau|0,T) \tag{51}$$

where $G(x,t|z,\tau)$ is the diffusion propagator

$$G(x,t|z,\tau) = \frac{1}{\sqrt{4\pi D(\rho)(t-\tau)}} \exp\left[-\frac{(x-z)^2}{4D(\rho)(t-\tau)}\right]. \tag{52}$$

These are the only two quantities required for simplifying the expression of $\langle X_T^2 \rangle$ in (45). Recalling that $q_1(x,0) = 0$ and $F_2 = 0$ in the quenched initial state we obtain

$$\frac{1}{2!}\langle X_T^2 \rangle_{\text{quenched}} = \frac{1}{\rho}\int_0^\infty q_1(x,T)dx - \frac{\sigma(\rho)}{2}\int_0^T dt \int_{-\infty}^\infty dx\,(\partial_x p_1(x,t))^2. \tag{53}$$

Using the solution for $q_1(x,t)$ and $p_1(x,t)$, the first integral yields

$$\frac{1}{\rho}\int_0^\infty q_1(x,T)dx = \frac{\psi(0,T)}{\rho}, \tag{54}$$

where we have taken into account that $\psi(x,T)$ vanishes at $x \to \infty$. The second integral in (53) becomes

$$\frac{\sigma(\rho)}{2}\int_0^T dt \int_{-\infty}^\infty dx\,[\partial_x p_1(x,t)]^2 = \frac{\psi(0,T)}{2\rho}, \tag{55}$$

where we have used $G(0,T|x,t) = R(x,t|0,T)$. Combining these results we reduce the expression for the variance to

$$\frac{1}{2!}\langle X_T^2 \rangle_{\text{quenched}} = \frac{\psi(0,T)}{2\rho} = \frac{\sigma(\rho)}{2\rho^2}\int_0^T dt \int_{-\infty}^\infty dx\, G(0,T|x,t)R(x,t|0,T)$$

Computing the integral we arrive at

$$\frac{1}{2!}\langle X_T^2 \rangle_{\text{quenched}} = \frac{\sigma(\rho)}{2\rho^2}\frac{\sqrt{T}}{\sqrt{2\pi D(\rho)}} \tag{56}$$

reported in the *Letter*.

*3.2. Annealed case*

Using the perturbative expansion (40)–(41) and the corresponding boundary conditions (13)–(14) we get

$$p_1(x,T) = \frac{\theta(x)}{\rho} \quad \text{and} \quad q_1(x,0) = \frac{\sigma(\rho)}{2D(\rho)}\left[p_1(x,0) - p_1(x,T)\right]. \tag{57}$$

Similarly, the variance (45) yields

$$\frac{1}{2!}\langle X_T^2 \rangle_{\text{annealed}} = \frac{1}{\rho}\int_0^\infty [q_1(x,T) - q_1(x,0)]\, dx - \frac{D(\rho)}{\sigma(\rho)}\int_{-\infty}^\infty dx\,(q_1(x,0))^2$$
$$- \frac{\sigma(\rho)}{2}\int_0^T dt \int_{-\infty}^\infty dx\,(\partial_x p_1(x,t))^2, \tag{58}$$

where we have used (44) and (46).

In the following, we determine the variance by drawing comparison with the quenched case. First, note that, in both cases the equation for $p_1(x,t)$ and the boundary condition on $p_1(x,T)$ are identical, leading to the same solution in (50). On the other hand, the equation (43) for $q_1(x,t)$ is same in both cases, although the boundary

condition is different. Considering that the equation is linear in $q_1(x,t)$, we write the solution in two parts

$$q_1(x,t) = q_I(x,t) + q_h(x,t), \tag{59}$$

where $q_I(x,t)$ is the solution of the inhomogeneous equation

$$\partial_t q_I - D(\rho)\partial_{xx} q_I = -\sigma(\rho)\partial_{xx} p_1 \quad \text{with} \quad q_I(x,0) = 0, \tag{60}$$

and $q_h(x,t)$ is the solution of the homogeneous equation

$$\partial_t q_h - D(\rho)\partial_{xx} q_h = 0 \quad \text{with} \quad q_h(x,0) = \frac{\sigma(\rho)}{2D(\rho)}\left[p_1(x,0) - p_1(x,T)\right]. \tag{61}$$

Comparing two initial states we notice that $q_I(x,t)$ is same as the $q_1(x,t)$ in the quenched case. Further, using (53) and (58) we conclude that

$$\frac{1}{2!}\langle X_T^2\rangle_{\text{annealed}} - \frac{1}{2!}\langle X_T^2\rangle_{\text{quenched}} = -\frac{D(\rho)}{\sigma(\rho)}\int_{-\infty}^{\infty} dx\, (q_h(x,0))^2$$
$$+ \frac{1}{\rho}\int_0^{\infty} dx\,[q_h(x,T) - q_h(x,0)]. \tag{62}$$

Thus the difference depends only on $q_h(x,t)$.

Further simplification comes from the identity

$$\frac{1}{\rho}\int_0^{\infty} dx\,[q_h(x,T) - q_h(x,0)] = \frac{2D(\rho)}{\sigma(\rho)}\int_{-\infty}^{\infty} dx\,(q_h(x,0))^2, \tag{63}$$

which can be proved by using (57) and noting that $\int_{-\infty}^{\infty} dx\, p_1(x,t) q_h(x,t)$ is a conserved quantity. Thus

$$\frac{1}{2!}\langle X_T^2\rangle_{\text{annealed}} - \frac{1}{2!}\langle X_T^2\rangle_{\text{quenched}} = \frac{D(\rho)}{\sigma(\rho)}\int_{-\infty}^{\infty} dx\,(q_h(x,0))^2$$
$$= \frac{\sigma(\rho)}{4D(\rho)}\int_{-\infty}^{\infty} dx\,[p_1(x,T) - p_1(x,0)]^2, \tag{64}$$

where in the last step we have used the boundary condition for $q_h(x,t)$ in (61).

The last integral can be performed using the solution (50) to yield

$$\frac{1}{2!}\langle X_T^2\rangle_{\text{annealed}} - \frac{1}{2!}\langle X_T^2\rangle_{\text{quenched}} = \frac{\sigma(\rho)}{2\rho^2}\left(\frac{\sqrt{2}-1}{\sqrt{2\pi}}\right)\sqrt{\frac{T}{D(\rho)}}. \tag{65}$$

Substituting the expression for the quenched case in (56) leads to the simple relation

$$\langle X_T^2\rangle_{\text{annealed}} = \sqrt{2}\,\langle X_T^2\rangle_{\text{quenched}}, \tag{66}$$

reported in the *Letter*.


	\end{center}
\end{widetext}

\end{document}